\documentclass[12pt,a4paper,leqno]{amsart}
\usepackage{latexsym, amsfonts, amsthm, amsmath, color, amssymb,
  graphicx, epsfig, float}
\usepackage[mathscr]{euscript}

\usepackage[bookmarks,bookmarksopen,colorlinks]{hyperref}
 \makeatletter
 \newif\ifmsbmloaded@

\input{scrload}

\title{Hidden Polynomial(s) Cryptosystems}
\author{Ilia Toli}
\address{Dipartimento di Matematica
  {\it Leonida Tonelli}\\ via F. Buonarroti 2,\newline 56127 Pisa\\
  Italy. {\tt toli@posso.dm.unipi.it} }
\date{}
\begin{document}
\begin{abstract}We propose variations of the class of hidden monomial
  cryptosystems in order to make it resistant to all known attacks. 
We use identities built upon a single bivariate polynomial equation with
  coefficients in a finite field. Indeed, it can
  be replaced by a ``small'' ideal, as well. Throughout, we set up
  probabilistic encryption protocols, too. The same ideas
  extend to digital signature algorithms, as well. Our schemes work as
  well on differential fields of positive characteristic, and
  elsewhere.\end{abstract}     
\maketitle
\section{Introduction}
This paper focuses on Hidden Monomial Cryptosystems, a class of
public key cryptosystems first proposed by Imai and
Matsumoto \cite{imai1}.  In this class, the
public key is a set of polynomial nonlinear equations. The private key
is the set of parameters that the user chooses to construct the equations.
Before we discuss our variation, we review
briefly a simplified version of the original cryptosystem, better
described in \cite{koblitz}. 
The characters  met throughout this paper are:
\begin{itemize}
\item Alice who wants to receive secure messages;
\item Bob who wants to send her secure messages;
\item Eve, the eavesdropper. \end{itemize}

Alice takes two finite fields $\mathbb{F}_q<\mathbb{K}$, $q$ a power of
$2$, and $\beta_1, \beta_2, \ldots , \beta_n$ a basis of
$\mathbb{K}$ as an $\mathbb{F}_q$-vector space. Next she takes $0<h<q^n$
such that $h=q^{\theta}+1$, and $gcd(h,q^n-1)=1$. Then she takes two
generic vectors ${\bf
  u}=(u_1,\ldots,u_n)$ and ${\bf v}=(v_1,\ldots,v_n)$ upon $\mathbb{F}_q$, and
puts\footnote{In this paper we reserve {\bf boldface}
   to the 
  elements of $\mathbb{K}$ thought as vectors upon $\mathbb{F}_q$ in
  the fixed private basis. They are considered vectors or field
  elements, as convenient, without further 
  notice. This shift in practice takes a Chinese Remainder Theorem. In
  order to avoid boring repetitions, {\it 
    Cryptosystem} and {\it Scheme} are used like synonyms.}: 
\begin{equation}{\bf   
  v=u}^{q^{\theta}} {\bf u}.\label{vuu}\end{equation} 

 The condition $gcd(h,q^n-1)=1$ is equivalent to requiring that the map ${\bf
  u}\longmapsto~{\bf u}^h$ on $\mathbb{K}$ is  ${\it
  1\!\!\leftrightarrow\!\!1}$; 
  its inverse  
  is the map ${\bf u}\longmapsto {\bf u}^{h'},$ where $h'$ is the
  inverse multiplicative of $h$ modulo $q^n-1$.

In addition, Alice chooses two secret affine transformations, i.e.,
two invertible matrices $A=\{A_{ij}\}$ and $B=\{B_{ij}\}$ with entries
in $\mathbb{F}_q$, and two constant vectors ${\bf c}=(c_1,\ldots,c_n)$
and ${\bf d}=(d_1,\ldots,d_n)$.

Now she sets:
\begin{equation}{\bf u}=A{\bf x+c}\qquad and \qquad {\bf v}=B{\bf
    y+d}.\label{aff}\end{equation} 

 Recall that the operation of raising to the
$q^k$-th power in $\mathbb{K}$ is an $\mathbb{F}_q$-linear
transformation.
Let $P^{(k)}=\{p_{ij}^{(k)}\}$ be the matrix of this
linear transformation in the basis $\beta_1, \beta_2, \ldots ,\beta_n$, i.e.:
\begin{equation} 
\beta_i^{q^k}=\sum_{j=1}^n p_{ij}^{(k)}\beta_j, \qquad
p_{ij}^{(k)}\in\mathbb{F}_q , \label{id1}
\end{equation}
for $1\leq i,k\leq n$. Alice also writes all products of basis elements
in terms of the basis, i.e.:
\begin{equation} 
\beta_i\beta_j=\sum_{\ell=1}^n m_{ij\ell}\beta_{\ell}, \qquad m_{ij\ell}\in
\mathbb{F}_q, 
\label{id2}\end{equation}
for each $1\leq i,j\leq n$. 
Now she expands the equation (\ref{vuu}). So she obtains a system of
equations, explicit in the $v$, and quadratic in the $u$. She uses now
her affine relations (\ref{aff}) to replace the $u,v$ by the
$x,y$. So she obtains $n$ equations, linear in the $y$, and of degree
$2$ in the $x$. Using linear algebra, she can get $n$ explicit
equations, one for each $y$ as polynomials of degree $2$ in the $x$.

Alice makes these equations public. Bob to send her a message $(x_1,
x_2, \ldots ,x_n)$, 
substitutes it into the public equations. So he obtains a linear system of
equations in the $y$. He solves it, and sends  ${\bf y}=(y_1,
y_2,\ldots,y_n)$ to Alice. 

To eavesdrop, Eve has to substitute
$(y_1,y_2, \ldots ,y_n)$ into the public equations, and solve the
nonlinear system of equations for the unknowns $x$.

When Alice receives {\bf y}, she decrypts:
\begin{eqnarray*}&y_1, y_2,\ldots,y_n&\\
&\Downarrow&\\
&{\bf v}=B{\bf y+d}&\\
&\Downarrow&\\
&{\bf v}=\sum v_i\beta_i &\\
&\Downarrow&\\
&{\bf u=v}^{h'}&\\
&\Downarrow&\\
&{\bf x}=A^{-1}({\bf u-c}).&
\end{eqnarray*}

In Eurocrypt $'88$ \cite{imai2}, Imai and Matsumoto proposed a digital
signature algorithm for their cryptosystem. 
At Crypto $'95$, Jacques Patarin \cite{Patarin95} showed how to break this
cryptosystem. He noticed that if one takes the equation  ${\bf
  v=u}^{q^{\theta}  +1}$, raises both sides on the $(q^{\theta}-~1)$-th
power, and multiplies both sides by ${\bf uv}$, he gets the equation ${\bf
  u v}^{q^{\theta}}={\bf u}^{q^{2\theta}} {\bf v}$ that
leads to equations in the $x$, $y$, linear in both sets of
variables. Essentially the equations do not suffice to identify uniquely
the message, but now even an exhaustive search will be
feasible. The system was definitively insecure and breakable, but its
ideas inspired a whole class of public key cryptosystems and digital
signatures based on structural identities for finite field operations
\cite{HFE, moh, koblitz, Patarin96, patarin96hidden, gou-pat1}.   

Actually, the security of this class lies on the difficulty of the
problem of solving systems of polynomial equations. This problem is
hard iff the equations are randomly chosen. All manipulations aim to
make equations seem like that. If they really were random, the problem
is hard to Alice, too. 

Our paper is organized as follows. In the next section we develop our
  own, new cryptosystem. Alice builds her public key by manipulations
  as above, starting from a certain bivariate polynomial. All of
  Alice's manipulations are meant to hide from Eve this polynomial. It
  is the most important part of the private key. Its knowledge reduces
  decryption to the practically easy problem of solving a single
  univariate polynomial.

In the third we discuss some security issues. There we explain that
practically all bivariate nonlinear
polynomials are good to us to give raise to a public key. This
plentitude of choices is an important security parameter.

In the fourth section we provide our cryptosystem with a digital
signature algorithm. 
In the fifth one we provide one more encryption protocol, now a
probabilistic one, in the sense that to the same cleartext correspond
zero, one, or more cyphertexts. 

In the sixth one we discuss some more variations. Essentially, we
replace the single bivariate polynomial by an ideal of a small size.

In the seventh section we mention what Shannon
  \cite{stinson} calls {\it
  Unconditionally Secure Cryptosystems.} Actually, this class of
  cryptosystems is considered an exclusive domain of private key
  cryptography. This is due mostly to the unhappy state of art of
  public key cryptography. 

In the eighth one we extend our constructions to differential fields
of positive characteristic. We hope they are the suitable environment
for unconditionally secure public key cryptosystems.
\section{A New Cryptosystem}
\subsection{Key Generation}
Alice chooses two finite fields
 $\mathbb{F}_q <\mathbb{K}$,  
 and a basis $\beta_1, \beta_2,\ldots, \beta_n $  of
 $\mathbb{K}$ as an  $\mathbb{F}_q$-vector space. Next she 
takes a generic (for now) randomly chosen bivariate polynomial:
\begin{equation}f(X,Y)=\sum_{ij}{{\bf a}_{ij}X^iY^j\label{poly1}}\end{equation}
in $\mathbb{K}[X,Y]$, such that she is able to find {\bf all} its roots in
$\mathbb{K}$ with respect to $X$; $\forall$ $Y \in \mathbb{K}$, if any. 
For the range of $i$ employed, this is nowadays considered a relatively
easy problem. Further, $f(X,Y)$ is subject to other few constraints, that
 we make clear at the opportune moment.

In transforming cleartext into ciphertext message, Alice will work
with two intermediate vectors, ${\bf u}=(u_1,\ldots,u_n)$ and ${\bf
  v}=(v_1,\ldots,v_n)$; ${\bf u, v \in \mathbb{K}}$. 
She sets: 
\begin{equation}
\sum_{ij}{{\bf a}_{ij}{\bf u}^i{\bf
      v}^j}=0. \label{poly} \end{equation}  
 
For ${\bf a}_{ij} \neq 0$, she sets somehow: 
 \begin{equation} 
i=\sum_{k=1}^{n_{i}} q^{\theta_{ik}},\qquad
j=\sum_{k=1}^{n_{j}} q^{\theta_{jk}}, 
\label{equal}\end{equation}
where $\theta_{ik}, \theta_{jk} n_{i}, n_j,\in\mathbb{N}_*$. 
Here {\it somehow} means that (\ref{equal}) {\bf need not} be the $q$-ary
representation of $i$, $j$. Indeed, there is no reason for it to be. We
allow to each $i$ both opportunities: to be or not to be. Doing so we
increase our choices, whence the random-looking of the public key. In
any fashion, what we are dealing with, are nothing but identities.

Next Alice substitutes the (\ref{equal}) to the exponents in
(\ref{poly}), obtaining:
\begin{equation}
\sum_{ij}({{\bf a}_{ij} exp({\bf u},{\sum_{k=1}^{n_i}
  q^{\theta_{ik}}}) exp({\bf
  v},{\sum_{k=1}^{n_0} 
  q^{\theta_{jk}}})})=0;
\end{equation} 
that is:
\begin{equation}
\sum_{ij}({{\bf a}_{ij} \prod_{k=1}^{n_i}{\bf u}^{
  q^{\theta_{ik}}}}\prod_{k=1}^{n_j}{\bf v}^{
  q^{\theta_{jk}}}) =0.
\label{prod}\end{equation}

{\bf Recall that the operation of raising to the
$q^k$-th power in $\mathbb{K}$ is an $\mathbb{F}_q$-linear
transformation.} 
Let $P^{(k)}=\{p_{\ell m}^{(k)}\}$ be the matrix of this
linear transformation in the basis $\beta_1, \beta_2, \ldots ,\beta_n$, i.e.:
\begin{equation} 
\beta_{i}^{q^k}=\sum_{j=1}^n p_{ij}^{(k)}\beta_j, \qquad
p_{ij}^{(k)}\in\mathbb{F}_q ; \label{id3}
\end{equation}
for $1\leq i,\,j\leq n$. Alice also writes all products of basis elements
in terms of the basis, i.e.:
\begin{equation} 
\beta_{i}\beta_j=\sum_{k=1}^n m_{ijk}\beta_{k}, \qquad
m_{ijk}\in\mathbb{F}_q; 
\label{id4}\end{equation}
for $1\leq i,\,j\leq n$. 

Now she  substitutes ${\bf u}=(u_1, u_2,\ldots,u_n)$, ${\bf a}_{ij}=(a_{ij1},
a_{ij2},\ldots,a_{ijn})$,
${\bf v}=(v_1,v_2,\ldots,v_n)$, and the
identities (\ref{id3}), (\ref{id4}) to (\ref{prod}), and
expands. So she 
obtains a system of $n$ equations of degree $t$ in
the $u$, $v$, where:
\begin{equation}t\ =\ max \  \{n_{i}+n_j\ \ :\ \
   {\bf a}_{ij}\neq 0\}.\label{set}\end{equation} 

Every term under $\Sigma$ in (\ref{equal}) contributes by one to the degree
in the $u$ of the polynomials.

Here we pause to give some constraints on the range of $i$, $j$ in
(\ref{poly}). The 
aim of this section is to generate a set of polynomials; linear in a
set of variables, and nonlinear in another one. For that purpose, we
relate (\ref{poly}) and (\ref{equal}): ${\bf a}_{ij}\neq 0
\Rightarrow$ $\{n_i>1$, $n_j=1\}$.

On the other side, the size of public key will be
$\mathcal{O}((2n)^{t+1})$. So, it grows polynomially with $n$, and
exponentially with $t$. Therefore, we are interested to keep $t$
rather modest, e.g., $t=2,3$ or so. So, we
have to choose $i$, $j$ in (\ref{poly1}), (\ref{equal})  in order to
keep $t$ under a forefixed bound.

Next, Alice chooses $A=\{A_{ij}\}, B=\{B_{ij}\}\in GL(\mathbb{F}_q)$,
${\bf c}, {\bf d}\in\mathbb{K}$, and sets: 
\begin{equation}
{\bf u}=A{\bf x+c},   \qquad {\bf v}=B{\bf y+d}, \label{matrix}
\end{equation}
where ${\bf x}=(x_1,x_2,\ldots,x_n)$, ${\bf y}=(y_1,y_2,\ldots,y_n)$ are
vectors of variables.

Now she substitutes  (\ref{matrix}) to the equations in the $u$,
$v$ above, and expands. So she  
obtains a system of $n$ equations of degree $t$ in the $x$, $y$;
linear in the $y$, and nonlinear in the $x$.

After the affine transformation, in each equation appear terms of each degree,
from zero to $t$; before not. This is its use; to shuffle terms coming
from different monomials of (\ref{prod}).

At this point, we are ready to define the cryptosystem. 
\subsection{The Protocol}With the notations adopted above, we define
the {\bf HPE 
  Cryptosystem} (Hidden Polynomial Equations) as the public
  key cryptosystem such that:
\begin{itemize}
\item The public key is:
\begin{itemize}\item The set of the polynomial
    equations in the $x$, $y$ as above;
\item The field $\mathbb{F}_q$;
\item The alphabet: a set of elements of $\mathbb{F}_q$.
\end{itemize}
\item The private key is: \begin{itemize}
\item The polynomial (\ref{poly1});
\item $A$, $B$, ${\bf c}$, ${\bf d}$ as in (\ref{matrix}); 
\item The identities (\ref{poly}) to (\ref{id4});
\item The field $\mathbb{K}$.
\end{itemize}
\item Encryption:\par Bob separates the cleartext $M$ by every $n$
letters. If needed, he
completes the last string with empty spaces. Next he takes an $n$-tuple
${\bf x}=(x_1,x_2,\ldots,x_n)$ of $M$, substitutes it to the $x$ in the 
public equations, solves with respect to the $y$, and sends ${\bf
  y}=(y_1,y_2,\ldots,y_n)$ to Alice. We assume here that the
solutions exist, and postpone the case they do not.  
\item  Decryption: \par Alice substitutes
  ${\bf v}=B^{-1}({\bf y-d})\in\mathbb{K}>\mathbb{F}_q$ in 
(\ref{poly}), and finds {\bf all} solutions within $\mathbb{K}$.  
There is at least one. Indeed, if ${\bf x}$ is Bob's cleartext, ${\bf
  u}$ as in (\ref{matrix}) is one. 
For each solution ${\bf u}$, she solves:
  \begin{equation}{\bf x}=A^{-1}({\bf u-c}),
  \label{expl}\end{equation}and represents all solutions in the basis
  $\beta_1, \beta_2,\ldots, \beta_n $. It takes a Chinese Remainder
  Theorem. With probability $\approx 1$, all 
results but one, Bob's $(x_1,x_2,\ldots,x_n)$, are gibberish, or even stretch
out of the alphabet.
\end{itemize} 
\subsection{Remarks}\subsubsection{}The risc of uncertain decryption
  is quite virtual. It equals
  the probability that another sensate combination of letters ${\bf
  x}$ satisfies (\ref{expl}) for any root ${\bf u}$ of (\ref{poly})
  for Bob's ${\bf y}$, besides the good one that always
  does. Afterwards, the undesired solution has to join well with the other
  parts of the decrypted message.
\subsubsection{}The main suspended question is that of existence of
  solutions. Well, Bob succeeds to encrypt a certain message {\bf x}
  iff Alice's equation (\ref{poly}) has solutions for {\bf u} as in
  (\ref{matrix}) for that {\bf x}. Alice's polynomial is a random
  one. It is a well-known fact from algebra that the
probability that a random polynomial of degree $m$ with coefficients upon a
field $\mathbb{F}_{q^n}$ has a root in it is about
$1-\frac{1}{e}\approx 63.2\%$ \cite{koblitz, marcus}. 
\label{remedy}
Now the remedy is probabilistic. Alice renders the alphabet public
with letters being sets of $\mathbb{F}_q$. Bob writes down a plaintext
and gives start to encryption. If he fails, he substitutes a letter of
the cleartext with another one of the same set, and retries.

After $s$ trials, the probability 
he does not succeed is $\approx \frac{1}{e^s}$; sufficiently small for
the algorithm to be trusted to succeed.
\subsubsection{}The other problem is that Alice may have to
distinguish the right solution among a great number of them. Here we
propose a first remedy. Her number of solution is bounded above by the
degree in $X$ of $f$. So, it is beter to her to keep this degree
moderate. Later in this paper in other settings, there will be other
remedies, too.

There are no bounds on the degree in $Y$. It can be taken
whatsoever huge.
\subsubsection{}Solving univariate polynomial equations is used by
  Pa\-ta\-rin, too \cite{patarin96hidden, Wolf:02:Thesis}. He takes a
  univariate polynomial:
  $$f(x)=\sum_{i,j}\beta_{ij}x^{q^{\theta_{ij}}+q^{\varphi_{ij}}}+
  \sum_i\alpha_ix^{q^{\xi_i}}+\mu_0,$$
and with manipulations like ours, both the same as Imai-Matsumoto
  \cite{imai1}, he gets his public key; a set of
  quadratic equations. He uses two
  affine transformations to shuffle the equations. We claim that the
  first one adds nothing to the security.

The bigger the degree of $f$ is, the more the public key resembles a
  randomly chosen set of quadratic equations. So, it is a security
  parameter.  On the other side, it slows down decryption, principally
  by adding a 
  lot of undesired solutions. To face that second problem, to the
  public key are added other, randomly chosen, equations. This is its
  {\it Achilles' heel}. It
  makes the public key overdefined, therefore subject to certain
  facilities to solve \cite{ckps}. So, it weakens the trapdoor
  problem.

We do not add equations to discard
undesired solutions. 
So, we are not subject to overdefined stuff. If in certain variations
we do add, we need to add less equations, however. 
We label {\it wrong solutions} those
  that after decrypted do not make sense, or stretch out of the
  alphabet. 

Afterall, all decrypted texts will howsoever be in a
  comprehensible language (to someone or some wedget). As $n$ grows,
  it is less possible to have more than one meaningful
  solution. Besides, any monkey solution that appears to Alice,
  appears to Eve, too. Furthermore, Eve may have more meaningful solutions.
If desired,
  other tests 
  can be introduced for that purpose. There is no need, however. The
  solutions, the good one and the bad ones, are very few; no more than $m$.

A big advantage
  of our settings is that we need a lower degree
  polynomial in $X$. So, we make the presence of
  undesired solutions virtual. Decryption is a pure
  linear algebra matter. 

What is most important, we have now a practically infinite range of choices of
$f$. This is not Patarin's case. There the choices are bounded below
because of being easy to attack cases, and above because of being impractical
to legitimate users.

The only few constraints we put on its monomials aim to:
\begin{itemize}
\item keep public key equations linear in the $y$; 
\item have less undesired solutions in decryption process;
\item keep the  size of public key moderate;
\item keep {\bf all} public key equations nonlinear in th $x$.
\end{itemize}\label{bivar}

 We can
  take the degree in {\bf y} unreasonably high. It 
  gives no trouble to us. It suffices that all the powers of {\bf y}
  that appear in the monomials of $f$ are powers of $q$, so the
  public equations come linear with respect to the $y$. 

A new facility now is that we can take lower degree in {\bf x},
as {\it multiple linear attack} does not anymore apply, hopingly.

The constraint that {\bf all} public key equations {\bf must} be
nonlinear in the $x$ is the only non-negotiable one. Indeed, if Alice
violates it, the trapdoor problem becomes fatally easy to Gr\"obner
techniques.

Back to the degree in the $y$ of the public key. Assume that the public
key equations are not linear in the $y$. Once Bob substitutes the
$x$ in the public equations, he now {\bf is not} challenged to solve a
nonlinear 
system of equations. He is only required to {\bf find one solution of
  it}. This can be done within polynomial time with respect to the
total degree of the system. Later we give settings to keep public key
nonlinear of modest degree in the $y$.

Each of such solutions (if any) is encryption to the same cleartext. So
we have set up a probabilistic encryption protocol. To a single cleartext
may correspond zero, one, or more ciphertexts.

So, in conclusion, Alice is allowed to take for the construction of
her public key {\bf any damned bivariate polynomial}. Indeed, we later
argue that $f$ can quite well be a multivariate polynomial. 

We hope this plentitude of choices is a spoil-sport to Eve.
\section{Security Issues}
Apparently, the only things Eve knows, are the system of public
equations, and the 
order of extension. By brute force, she has to take
$(y_1,y_2,\ldots,y_n)$, to substitute it in the public key equations, to
solve in $\mathbb{Z}$, or maybe $\mathbb{Z}[\alpha]$, and to take the sensate
solution. Almost surely, 
there is only one good solution among those that she finds.
 She has to find it among $t^n$ of them. However, the  main difficulty
 to her is just 
 solving the system. Supposedly, it will pass through the complete
 computation 
 of Gr\"obner basis. It is a well-known hard problem. The
 complexity of computations upon a field grows at most twice
 exponentially with respect to the 
 number of variables, and in the average case, exponentially. 

So, it is better to take  
$n$ huge. This diminishes the probability that Alice confuses decryption,
however close to zero, and, what is most important, this renders Eve's
task harder. 

 Alice and Bob will have to solve sets of bigger systems of
 linear equations, and face Chinese Remainder Theorem for bigger $n$.

There exist well-known facilities \cite{ckps} to solve overdefined systems of
equations. Unlike most of the rest, our public key is irrendundant, so
it is not subject to such facilities.

Now, by exhaustive search we mean that Eve substitutes the ${\bf y}$ in the
public equations, and tries to solve it by substituting values to
${\bf x}$.

If we have $d$ letters each of them being represented by a single
element of $\mathbb{F}_q$, the complexity of an exhaustive search is
$\mathcal{O}(d^n)$. It is easy for Alice to render exhaustive search
more cumbersome than 
Gr\"obner attack. The last one seems to be the only choice to Eve.

We did not find any {\it Known Cleartext Attack} to our cryptosystem.

Eve may engineer {\it
  cleartext$\,\leftrightarrow\,$ciphertext 
  analyses}, seeking for invariants or regularities there, helpful for an
attack \cite{patarin96hidden}. All the identities we use, mean to
tousle any such regularity, 
and to disguise from Eve any hint on $i$, $j$, and on the entries of 
$A$, $B$, ${\bf c}$, ${\bf d}$, and the ${\bf a}_{ij}$; that she may
  use for such an attack.  

The complexity of the trapdoor problem is $\mathcal{O}(t^n)$,
the size of public key $\mathcal{O}(n^{t+1})$. This fully suggests the
values of parameters. $n=100$, $t=2,3,4$ would be quite good choices.

Obviously, infinitely many bivariate polynomials give raise to the same public
key. Indeed, fixed the ground field, the degree of extension $n$, and
the degree of public key equations, we have a finite number of public
keys. On the other hand, there are infinitely many bivariate polynomials that
can be used like private keys. 

On how does it happen, nothing is known. If ever found, any such 
regularity will only weaken the trapdoor problem.

\section{A Digital Signature Algorithm}\label{sign}
Assume that we are publicly given a set of hash functions that send
cleartexts to strings of integers of fixed length $n_B$. For the only
purpose of signing messages for Alice, Bob builds a cryptosystem as above
with $q_B$ 
prime, and $[\mathbb{K}_B:~\mathbb{F}_{q_B}]=n_B$.
He to sign a message $M$:
\begin{itemize}
\item calculates
$H(M)=(y_1,y_2,\ldots,y_{n_B})={\bf y}\in \mathbb{K}_B $; 
\item finds one solution (if any; otherwise, see section
  \ref{remedy}.) {\bf u} of
  $f_B({\bf u})={\bf y}$ in $\mathbb{K}_B$.
\item calculates ${\bf x}=B^{-1}({\bf u-c}_B)$;
\item appends ${\bf x}=(x_1,x_2,\ldots,x_{n_B})$ to $M$, encrypts,
 and sends it 
 to Alice.  $(x_1,x_2,\ldots,x_{n_B})$ is a signature to $M$.\end{itemize}  
 
To authenticate, Alice first decrypts, then she:
\begin{itemize}
\item calculates $H(M)=(y_1,y_2,\ldots,y_{n_B})$; 
\item substitutes $(x_1,x_2,\ldots,x_{n_B})$, $(y_1,y_2,\ldots,y_{n_B})$ to
  Bob's public equations; 
\item so she gets an $n_B$-tuple of integers. If they all reduce to
  zero modulo $q_B$,  she accepts the message; otherwise she
knows that Eve has been causing trouble.
\end{itemize}

If Eve tries to impersonate Bob and send to Alice her own message with hash
value ${\bf y}=(y_1,y_2,\ldots,y_{n_B})$, then to find a signature
$(x_1,x_2,\ldots,x_{n_B})$, she may try to find one solution of Bob's system
of equations for {\bf y}.
We trust on the hardness of this problem for the security of
authentication.

\section{A Probabilistic Encryption Protocol}
With the ideas described above, we are going to set up now a
probabilistic protocol such that only the legitimate users can send
messages to which-another. Mean, the message is meaningful iff there
are no intruders. Its being meaningful is the signature itself. 

Here is the shortest possible description. Let $F_A$ and $F_B$ be
Alice's and Bob's public keys functions respectively, where $n_A=n_B$. To send
a message {\bf x} to Alice, Bob sends her a random (this randomness is
the probabilistic pattern) element of
$F_A(F^{-1}_B({\bf x}))$,
that she can decrypt by calculating 
$F_B(F^{-1}_A(F_A(F^{-1}_B({\bf x}))))$. So if $F_A(F^{-1}_B({\bf
  x}))\neq \emptyset$. Otherwise, the approach is probabilistic, as in
the previous section.

Here is the extended description. Each (English, e.g.)
letter (or some of them, only) is represented by a set of few
(two, e.g.) elements of the field, or 
strings of them. For ease of explanation, Bob's public equations are
linear in the $x$, and of higher degree in the $z$.

Bob writes down the cleartext ${\bf  x}$ and finds one
    solution of:\begin{equation} {\bf x}={\bf b}_r{\bf  z}^r+{\bf
    b}_{r-1}{\bf 
    z}^{r-1}+\cdots+{\bf b}_0 .\label{polyB}\end{equation}  

If there are no solutions, Bob changes a 
representant of a letter, and retries. Probability issues are discussed
in the previous section.

Now Bob takes the solution {\bf z} of (\ref{polyB}), and applies:
\begin{equation}{\bf y'}=B^{-1}({\bf z-c}_B).
\label{explB}\end{equation}

Next he takes ${\bf y'}$, substitutes in Alice's public
equations. So he obtains a tuple {\bf y}, that he sends to Alice. This
is the ciphertext. 

Each of other solutions of (\ref{polyB}) give
raise to other encryptions of the same cleartext. 

Alice now to decrypt, solves her equation for {\bf y} within her
field $\mathbb{K}$. There is at least one solution. Next she applies
her inverse affine transformation to all (few)
solutions, and substitutes them all on Bob's public equations. Of that
procedure all, Alice now discards all meaningless solutions, and takes the
meaningful one. 

What is the trapdoor problem now?
Well, on authentication matter, nothing new. Eve has the same chances
to forge here that she had before. Recall that this kin of signatures
is already best with respect to the other ones.

On security, instead, there is a very good improvement. By brute
force, Eve has to take the 
ciphertext, substitute on Alice's public key, find all solutions, and
substitute them all on Bob's public key; then take the sensate
ones. This is worse than exhaustive search of previous
cryptosystems. 

Now, what does here really mean {\it exhaustive search}? Eve now has
to search through all the elements of the common public ground field,
not just through all the alphabet. So, opting for this protocol, we
can put a lot of constraints on alphabet, 
in order to discard far easier the undesired solutions, without
rendering the public key overdefined.

She sets up such $n$-tuples, checks whether
they are solutions of Alice's public key for Bob's ciphertext
{\bf y} substituted to the variables $y$. If yes, she substitutes to
Bob's public key, and checks whether does it make sense.

What can {\it linear multiple attack} or {\it quadratic attack}
\cite{patarin96hidden} do in these new settings?

Apart all, we save space and calculi. We do not need any more the
calculi and space of signature. 

This protocol can be used for multiple encryption, too.

Let us suppose that the letters are strings of a fixed length. Well,
here Alice can impose that not all strings are letters. So, in
decryption she discards a priori the solutions that contain
non-letters. Doing so, she actually has a single good solution of her
polynomial, and saves herself the effort of appealing to other
tricks. In all the other schemes throughout, such a trick fatally
weakens the exhaustive search.
\section{Hidden Ideal Equations}Instead of a single bivariate polynomial,
Alice may choose to employ an ideal of a very modest size. She separates
the variables she
employs into two sets, $\{X_i\}$, $\{Y_j\}$; one for encryption, one
for decryption. She may decide to leave one of the equations employed
of higher degree in the $\{Y_j\}$ after manipulations, so she gives raise to a
probabilistic encryption protocol.
Alice's parameters are: 
\begin{itemize}
\item $n=[\mathbb{K}:\mathbb{F}_q]$;
\item the number $s_1$, $s_2$ of variables $\{X_i\}$, $\{Y_j\}$, respectively;
\item the number $r$ of private equations.
\end{itemize}

So, the number of public key equations is $n\cdot r$. The number of the
variables $x_{ij}$ is $n\cdot s_1$, and that of the $y_{kl}$ is
$n\cdot s_2$.

Alice's number of variables, the $\{X_i\}$, is insignificant so far, so she is
supposed to be able to appeal to Gr\"obner stuff in order to solve her
system of equations within the field of coefficients for Bob's
$\{Y_j\}$. 

What is most important here and throughout, if 
Bob succeeds to encrypt, Alice does always succeed to decrypt. 

For ease of treatment, assume now that Alice does not apply affine
transformations to her variables. Bob fails encryption for a certain
cleartext $(X_1,\dots X_{s_1})$ iff Alice's private ideal has no solutions
in the $Y$ for such an $(X_1,\dots X_{s_1})$. Alice's private ideal is a
random one. If she takes $r\leq s_2$, the probability that it has no
solutions is $\approx 0$, and $\approx 1$ for $r> s_2$. So, it
suffices that Alice takes $r\leq s_2$. The critical cases that
may supervene are faced simply changing alphabet.

With slight changes, this reasoning holds in the case that Alice
applies affine transformations, too. 

The real problem is indeed that the solutions to Alice may be too many; and in
any case finitely many, as the base field is finite. The best remedy
to that is that Alice takes $r=s_1$. So, the ideal that she obtains
after substitution of Bob's ciphertext is zerodimensional (quite easy
to cause it happen), and the number of solutions is bounded 
above by the total degree of the system. So, she can contain the
number of solutions by taking the total degree in the $\{X_i\}$
modest, and however each of them nonlinear. 

Alice can take all equations of
very low degree in the $X$, and then transform that basis of the ideal
they generate to another one of very high degrees in the $X$. So she
has a low Bezout number of the ideal, and higher degrees in the $X$,
and transformations as above can take place.
If she takes the first basis linear, the number of solutions of her
equations reduce to one: Bob's cleartext.

As soon as $r>s_1$, the public key becomes overdefined.

Alice applies a permutation to the equations and a renumeration to the
variables before publishing her key, so Eve does not know how are they
related. She may apply 
affine transformations, or may not, or may apply to only some of the
$X_i$, $Y_j$; at her discretion.

If $s_1< s_2$, the size of the ciphertext is
bigger than that of cleartext, and nothing else wrong. By this case,
encryption is practically always probabilistic. Indeed, even when the
equations are linear with respect to the $y_{kl}$, since there are more
variables than equations, the solutions exist, and are not unique.

Actually, Alice can take $s_2$ rather huge. She may choose to
manipulate some of the $Y_j$ within a subfield of $\mathbb{K}$, rather than
within $\mathbb{K}$. Doing so, she allows herself a big $s_2$, and a
contained size of the ciphertext. The number of the variables $y_{kl}$
now is no more $n\cdot s_2$.
\subsection{}Now the size of the public key is
$\mathcal{O}(s_1(n)^{t+1})$, and the complexity of the
trapdoor problem is $\mathcal{O}(t^{n\cdot s_1})$.

It is true that throughout the size of public key grows polynomially with
$n$, but before $n$ becomes interesting, the public key is already
quite cumbersome. 
So, opting for the choices of this section we have reasonable security with
much smaller values of $n$. $n=20$, or so, actually are quite good. We
are allowed some more values of $t$, too. 

\subsection{}There exist classes of ideals called {\it with doubly
  exponential ideal membership property} \cite{swanson}. These are the
  ideals for 
  which the calculus of a Gr\"obner basis cannot be done within
  exponential time on the number of variables, i.e., it can be done
  within doubly exponential time on the number of variables. It is very
  interesting to know whether can we employ them in some fashion in
  this class of cryptosystems. In any fashion, this is the theoretical
  limit for employing solving of polynomial systems of equations in
  public key cryptography.

\section{Some Considerations}
The idea of public key cryptography was
first proposed by Diffie and Hellman \cite{pkc}. Since then, it has
seen several vicissitudes \cite{odlyzko}.  

A trapdoor function is a map from cleartext units to ciphertext
units that can be feasibly computed by anyone having the
public key, but whose inverse function cannot be
computed without knowledge of the private key:\begin{itemize}
\item either because (at present, publicly)
  there is no theory to do it; 

\item or the theory exists, but the amount of calculations is
  deterring.\end{itemize}   

Cryptosystems with trapdoor problems of
the first kin are what Shannon \cite{stinson} calls {\it
  Unconditionally Secure Cryptosystems}. 

Actually, the aim is to make trapdoor problems be equivalent to
  time-honoured hard 
  mathematical problems. However, being of a problem hard or
  undecidable implies 
  nothing about the security of the cryptosystem \cite{barkee, odlyzko}. 
Recall that of all schemes ever invented, only two of
  them, $RSA$ \cite{rsa} and {\it ECDL} \cite{koblitz},
  are going to be broken (or, at least, are going to become
  impractical) by solving the hard problems they lie upon. The rest
  of them have been broken with theories 
  of no use to solve their hard problem. So, once
  more, it may happen 
  to be proved that solving systems of differential\&integral equations
  is undecidable, nevertheless several cryptosystems
  built upon them may be easy to break rather than secure.

The author is very fond of the idea of public key cryptography, and
believes howsoever in new developments that will make it fully suffice
for all purposes.

Actually, one tendency is that of investigating {\it poor
  structures}, mean, structures with less operations, like groups,
semigroups with cryptosystems upon the {\it word problem}
  \cite{anshel, yamamura, hughes}. Yamamura's paper \cite{yamamura}
  can be considered pioneering on secure
  schemes. Unfortunately, its scheme is still uneffective.
 
William Sit and the author are investigating cryptosystems upon
other algebraic structures. We are investigating among other things whether
is it possible to build effective secure schemes upon
differential fields of positive characteristic. We
hope that cryptography will arouse new interests on differential and
universal algebra, too, as it did in number theory and arithmetic
geometry. One reason of optimism is that in universal algebra one can
go on further with new structures and hard or undecidable problems
forever. Until now we have appealed 
to only the unary and binary arithmetic operations.
\section{Generalizations on Differential Fields}
Differential algebra is born principally due to the efforts of Ritt
\cite{ritt} to handle differential equations by means of
algebra. Actually, a differential field is a field with a set of unary
operations $'$ called derivatives that replace an element of the field
with another one such that $(a+b)'=a'+b'$ and $(ab)'=ab'+a'b$.

Good references in the topic are \cite{kolchin, sit2, ritt, sadik,
  kaplansky}. Kaplansky's book is probably the best introduction in
  the topic.

It is possible\footnote{Most of considerations given in this section are
  suggestions of professor Sit through private communications.} to
  generalize the schemes given throughout using
differential polynomials instead of (\ref{poly1}). Take 
$\mathbb{K}$ to be a finite
differential field extension of a differential field
$\mathbb{F}$ of positive characteristic\footnote{In zero
  characteristic numerical analysis tools seriously affect security,
  or at least constrain us to more careful choices. We shall
  not dwell on this topic here.}.
Any such $\mathbb{K}$ is defined by a system of linear homogeneous
differential equations, and there are structural constants defining
the operations for the derivations (one matrix for each derivation),
as well for multiplication. 

One can now replace (\ref{poly1}) with a
differential polynomial. The scheme works
verbatim. One can take (\ref{poly1}) to be of higher order and degree,
that is ok too, just like the algebraic case. 
Euler, Clairaut, or any of other well-studied classes of  equations,
or their compositions; each of them fully suffice.

The techniques described above for polynomials, if
applied to differential polynomials, will definitely make it much harder
to attack any protocol developed. Any affine transformation (by this is
meant a linear combination of the differential indeterminates with
not-necessarily constant coefficients, and this linear combination is
then substituted  {\it differentially}  in place of the differential
indeterminates) will not only even out the degrees, but also the orders
of the various partials, and making the resulting differential
polynomial very dense. 

However, there is one thing to caution about:
any time one specifies these structural matrices, they have to satisfy
compatibility equations. In the algebraic case, it is the relations
between $P^k=\{{p_{ij}}^{(k)}\}$ in (\ref{id3}) and
$M_{\ell}=\{m_{ij\ell}\}$ in (\ref{id4}). The $P^k$ are simply determined
uniquely by $M_{\ell}$, given the choices implicitely defined in (\ref{id4}).

It is very interesting to know in the algebraic case whether the
system of equations Alice obtains is invariant under a change of
basis, all other settings being equal. There is probably some group of
matrices in $GL(n, q)$ that can do that. Such a knowledge may be used to
build attacks to all schemes of $HFE$ class.

In the differential case there is a similar action called Loewy
action, or the gauge transformation. For ordinary differential
equations, two matrices $A$, $B$ are Loewy similar if there is an
invertible matrix $K$ such that $A=\delta K\cdot
K^{-1}+KBK^{-1}$. Using this action, one can classify the different
differential vector space structures of a finite dimensional vector
space. There is also a cyclic vector algorithm to find a special basis,
so that the differential linear system defining the vector space
becomes equivalent to a single linear $ODE$. 

If no other problems arise for the differential
algebraic schemes, there is however
one caution more for them to be unconditionally secure. We have to avoid the
exhaustive search. For that, Alice has to publish a finite alphabet
where each letter is represented by an infinite set, disjoint sets for
different letters. This is possible in differential fields, as
they are infinite. Alice renders the sets public parametrically, as
differential algebraic functions of elements of the base differential
field, and parameters, e.g., in $\mathbb{Z}$. Bob
chooses a letter, gives random values to parameters, obtains one
representant of the letter, and proceeds as above. In any case, if
$\mu$ is the order of public equations, any two elements $\Xi$, $\Theta
\in \mathbb{F}$ such that $(\Xi - \Theta)^{(\mu )}=0$ must represent
the same letter, if any. 

The main care for Alice is that the public key
equations must not fall into tractable classes by well-known means,
such as linear algebra. 

In the algebraic case such constructions do not make sense. Eve can
anyway appeal to Gr\"obner attack. Besides, in any fashion 
such data enable her to guess $q$.

The size of the public key now is actually $\mathcal{O}(n^{to+1})$,
where $o$ is the order of public key equations. Quite
explosive. However, a first tool to contain it is the low
characteristic of the field. So, we see a lot of monomials reduce to
zero. The best consolation is that we do not have to go far away with
parameters. The trapdoor problem is simply undecidable.
$n=20$ would fully suffice. Such a value is needed
  more in order to avoid uncertain decryption, however less probable in
  differential fields, as the range of solutions is infinite, than for growing 
  security.  Besides, if there was found some attack for  the $HDPE$
(Hidden Differential Polynomial Equations) scheme, it will work better
with $HPE$. As of now, $HDPE$ trapdoor problem seems undecidable, and the
scheme effective. The author is working to come up with concrete
examples of this kind of cryptosystems. Unfortunately, 
everything in the topic is still handmade, and therefore rather time-consuming.
\subsection*{Acknowledgments.}
The author would like to thank Don Coppersmith, Patrizia 
Gianni, Teo Mora, Massimiliano Sala, and Barry Trager for
many suggestions and fruitful discussions. The author is particularly
indebted to William Sit for several comments and improvements on earlier
drafts, and to his advisor, Carlo Traverso.

\addcontentsline{toc}{section}{Bibliography}
\bibliographystyle{alpha}
\bibliography{biblio}
\nocite{HFE, Patarin95, gathen, odlyzko, barkee, koblitz,
  marcus, moh, imai1, imai2, sit,  patarin96hidden, pkc, sadik,
  kolchin, sit2, ritt, hughes, anshel, yamamura, gathen, stinson,
  ckps, patarin96hidden, Wolf:02:Thesis, menezes, swanson}

\end{document}